# Towards an overall astrometric error budget with MICADO-MCAO

G. Rodeghiero[1*], J.-U. Pott[1], D. Massari[2,3], M. Fabricius[4], V. Garrel[4], H. Riechert[1], R. Davies[4], C. Arcidiacono[5], M. Patti[5], F. Cortecchia[5], G. Fiorentino[5], M. Lombini[5]

[1]Max Planck for Astronomy, Königstuhl 17, 69117 Heidelberg, Germany
[2]Kapteyn Astronomical Institute, University of Groningen, PO Box 800, 9700 AV Groningen, The Netherlands
[3]Leiden Observatory, Leiden University, P.O. Box 9513, 2300 RA Leiden, The Netherlands
[4]Max Planck for Extraterrestrial Phyiscs, Gießenbachstraße, 85741, Garching, Germany
[5]INAF Osservatorio Astronomico di Bologna, Via Gobetti 93/3, 40129 Bologna, Italy

**ABSTRACT**

MICADO is the Multi-AO Imaging Camera for Deep Observations, and it will be one of the first light instruments of the Extremely Large Telescope (ELT). Doing high precision multi-object differential astrometry behind ELT is particularly effective given the increased flux and small diffraction limit. Thanks to its robust design with fixed mirrors and a cryogenic environment, MICADO aims to provide 50 µas absolute differential astrometry (measure star-to-star distances in absolute µas units) over a 53" FoV in the range 1.2-2.5 µm. Tackling high precision astrometry over large FoV requires Multi Conjugate Adaptive Optics (MCAO) and an accurate distortion calibration. The MICADO control scheme relies on the separate calibration of the ELT, MAORY and MICADO systematics and distortions, to ensure the best disentanglement and correction of all the contributions. From a system perspective, we are developing an astrometric error budget supported by optical simulations to assess the impact of the main astrometric errors induced by the telescope and its optical tolerances, the MCAO distortions and the opto-mechanical errors between internal optics of ELT, MAORY and MICADO. The development of an overall astrometric error budget will pave the road to an efficient calibration strategy complementing the design of the MICADO calibration unit. At the focus of this work are a number of opto-mechanical error terms which have particular relevance for MICADO astrometry applications, and interface to the MCAO design.

**Keywords:** astrometry, distortion, tolerances, calibration.

## 1. INTRODUCTION

The Multi-AO Imaging Camera for Deep Observations (MICADO) is targeted to be the first light instrument of the Extremely Large Telescope (ELT). The primary goal and observing mode of MICADO is imaging, with a focus on differential astrometry with an accuracy of ~50 µas; the instrument provides also a spectroscopic mode with a resolution R~8000. The observing window of the instrument is 0.8-2.4 µm and when assisted by the Multi-conjugate Adaptive Optics RelaY (MAORY) it has a corrected FoV of 53" [1]. 50 µas differential astrometry is an ambitious goal and it requires careful estimations, control and calibration of all the systematics. This work concentrates only on the opto-mechanical tolerances of the ELT, MAORY and MICADO optics, accounting for quasi-static and dynamical drifts. The opto-mechanical tolerances are only one aspect of the full astrometric error budget [2], but they are particularly important in the perspective of scouting the worst offenders elements in the optical train and developing a suitable calibration plan for the observations. The method used in the current work is based on a selective sensitivity analysis to tolerances boosted by a Monte Carlo simulation approach. The core results are presented in Section 3 and 4.

---

*Corresponding author: rodeghiero@mpia.de; phone +49-(0)6221-528-258

## 2. MONTECARLO & TOLERANCE STUDY APPROACH

The approach adopted in current study is based on Monte Carlo (MC) simulations combined with a tolerance study for the different optics of the three main systems ELT, MAORY and MICADO. The software used is based on the so-called ZOS API libraries that allow to control and launch simulations in Zemax-OpticStudio [3] from a Matlab script/environment. Figure 1 shows the scheme of a $n^{th}$ MC simulation: (i) a certain mirror is subjected to a random tolerances state ($\Delta x, \Delta y, \Delta z, \Delta\theta x, \Delta\theta y$) (ii) the wavefront errors (WFE) are extracted and parameterized by the first 37 Zernike terms in Noll notation [4] (iii) the shape of M4 is modified to minimize the residual WFE by 37 Zernike terms, the telescope is refocused with two compensators: back focal distance and M3 position, and the field is steered by M5; (iiii) the geometric distortion at the telescope/instrument focal plane is extracted and fitted with a $n^{th}$ order polynomial.

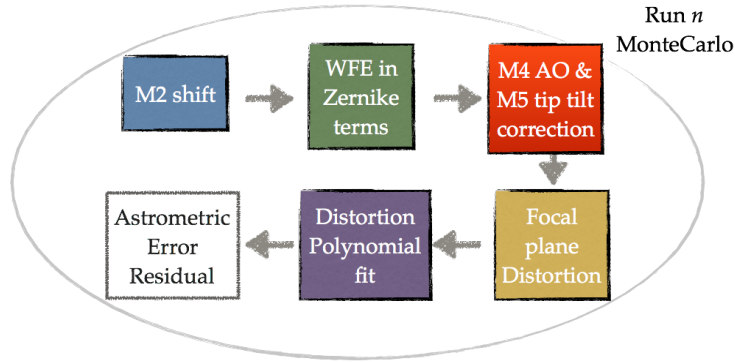

**Figure 1** Sequence of operations done by the simulation within a single MC realization. The whole calculation (ray tracing included) is implemented via Matlab interface to Zemax-OpticStudio.

To accurately disentangle the effect from each optical element, for each mirror of the ELT, 20 MC random realizations are produced, while for MICADO and MAORY we produce 20 MC realizations with all the optics perturbed simultaneously within the instrument.

The simulation has three main outputs: plate scale variation wrt. nominal design, exit pupil motion (either of the telescope or of the instrument) induced by the tolerances, and astrometric $RMS_x$ & $RMS_y$ residuals pre and post fit for $1^{st}$, $3^{rd}$ and $5^{th}$ order polynomials over the whole Field of View (FoV).

The tolerances ranges of certain optical elements have been assumed from the authors due to the lack of confirmed values.

## 3. MAIN RESULTS FROM SIMULATIONS

### 3.1 ELT tolerances and distortion

The range of tolerances for ELT M2 and M3 have been discussed by [5] and [6] and although these numbers could slightly vary from mirror to mirror, we assume the same values also for M4 and M5. The assumption is driven by the comparable size of the mirrors and by the interest of assessing the response to an equal-amplitude perturbation:

$$(\Delta x, \Delta y, \Delta z) \rightarrow \pm\, 0.1 \text{ mm}$$
$$(\Delta\theta x, \Delta\theta y) \rightarrow \pm\, 0.01 \text{ deg} \quad (1)$$

Figure 2 (left) shows the effect of tolerances (Eq. 1) in terms of telescope Plate Scale (PS) variations and exit pupil motion. The worst offender for PS variations is M2 that is highly sensitive to axial offsets wrt. M1: $\Delta z_{M2} = 0.1$ mm leads to $\Delta PS = 0.1\%$, $\Delta z_{M2} = 1$ mm leads to $\Delta PS = 1\%$. All the other mirrors (M3, M4, M5) induce very negligible PS variation in comparison with M2. M3 and M5 play instead an important role in the exit pupil motion (Figure 2, right), inducing shift of ~ 0.5 % of the exit pupil diameter; the exit pupil is located between M4 and M5. The amplitude of the optical distortions introduced by the ELT optics tolerances before the polynomial fit is shown in Figure 2 (bottom): M2 and M3, being powered optics give origin to the largest distortions and with a wide spread of values, when subjected to tolerances. Once the operations (i), (ii), (iii), described in Section 2 are performed in the simulation, the distortion pattern at the telescope focal plane is extracted to estimate the impact of the opto-mechanical tolerances on the astrometric precision.

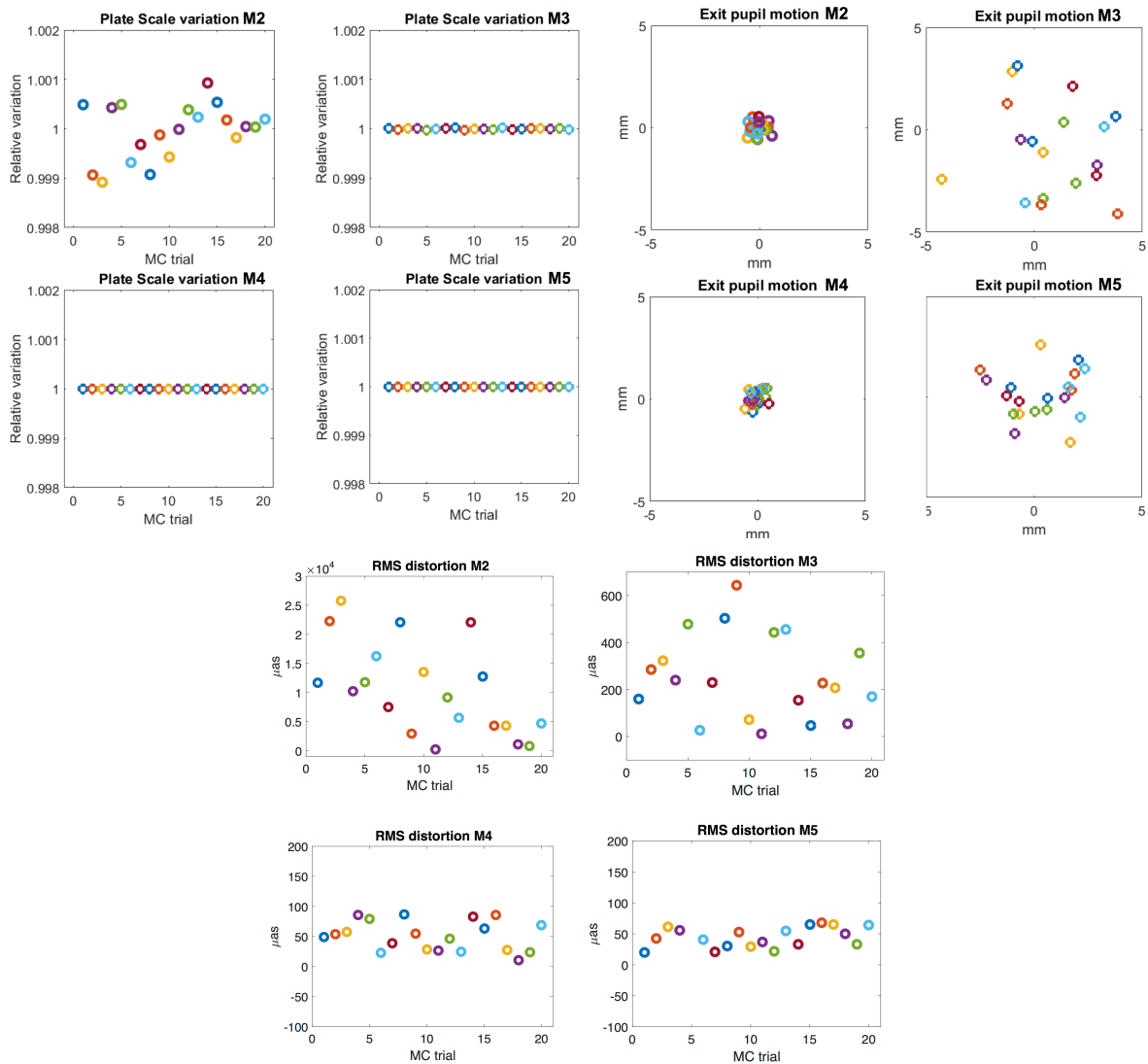

**Figure 2 Left:** Plate scale variation for ELT M2, M3, M4 and M5 tolerances within 20 MC simulations (color circles). The offset of M2 wrt M1 induces the largest plate scale variations. **Right:** exit pupil motion induced by tolerances on ELT M2, M3, M4 and M5: the major contributors are M3 and M5. **Bottom:** RMS distortion over the full FoV for the 20 MC simulations induced by the tolerances on ELT optics; M2 and M3 are the worst offenders being powered optics, M4 and M5 being flat mirrors have a lower impact on distortion production.

The distortion pattern at the ELT Focal Plane (FP) is sampled with an equally spaced grid of 144 points for all the MC realizations and the positions of the image points are fitted to the grid points obtained from the nominal ELT design. The latter contains a certain level of intrinsic optical distortion of the telescope nominal configuration, but it represents reference grid for our study. The polynomial fit expression Eq. 2 and Eq. 3 is the same used by [7]:

$$U = A_1 + A_2 X + A_3 Y + A_4 X^2 + A_5 XY + A_6 Y^2 + ... + A_{21} Y^5 \quad (2)$$

$$V = B_1 + B_2 X + B_3 Y + B_4 X^2 + B_5 XY + B_6 Y^2 + ... + B_{21} Y^5 \quad (3)$$

The U and V coordinates represent the grid points of the nominal ELT distortion pattern, while the X and Y coordinates are the points from the distortion pattern of a certain MC realization. The polynomial fit is performed for $1^{st}$, $3^{rd}$ and $5^{th}$ order. The $1^{st}$ order polynomial accounts for relative translation, rotation and plate scale variations between different distortion patterns, while the $3^{rd}$ and $5^{th}$ order polynomials describe higher order distortions. Figure 3 shows the post-fit RMS residuals of the whole grid (i.e. the whole FP) for 20 MC realizations of the tolerance study on each separate ELT mirror. The primary mirror of ELT being at the entrance pupil of the system does produce neither field-differential aberrations nor astrometric errors over the FoV so it is not discussed in this context. The astrometric systematics induced by M1 relate mainly to the decrease of Strehl ratio caused by the phasing errors of M1 segments. The other ELT mirrors are not in the entrance/exit pupil of the telescope and can cause distortion at different levels; the plots in Figure 3 shows the RMS residuals of the MC simulations distortion grids wrt the nominal ELT distortion for M2, M3, M4, M5 separately. The RMS residuals are given in comparison to the post-processing, final astrometric precision requirement for MICADO within a single epoch ($\sigma \sim 50$ µas). As the reader can observe, already the $1^{st}$ order fit leads to small astrometric residuals below the MICADO requirement, meaning that predominant ELT distortions are caused by plate scale variations. The $3^{rd}$ fit further decreases the RMS residuals leading to a compact cloud of RMS points with a centroid around 12-13 µas for all the mirrors, while higher order polynomials like e.g. $5^{th}$ order does not bring any additional improvement.

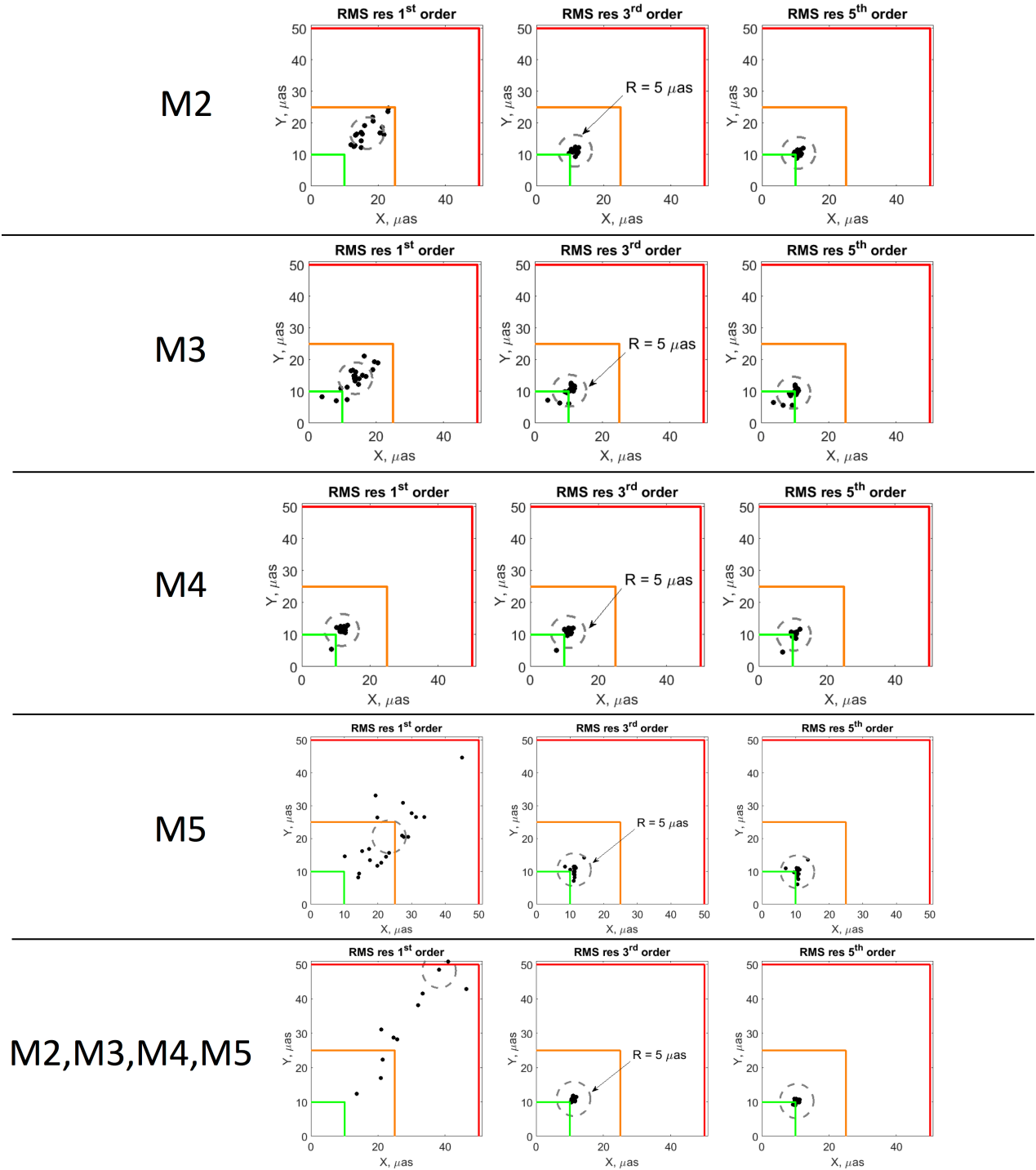

**Figure 3** RMS residual (post-fit) over the MICADO FoV between the n[th] MC grid distortion and the ELT nominal one after tolerancing the position and tilt of M2 (top to bottom), M3, M4 and M5. For each MC the RMS residuals are calculated after 1[st], 3[rd] and 5[th] order polynomial fit following equation (2) and (3). The 3[rd] fit removes the majority of the distortion and the 5[th] fit does not bring any significant further improvement (neither RMS lowering nor increased residuals compactness).

## 3.2 MAORY tolerances and distortion

The Multi-conjugate Adaptive Optics RelaY (MAORY) module will provide MICADO with the MCAO correction over the full instrument FoV (53"). MAORY will be equipped with two post focal DMs optically conjugated at 4 km and 12.7 km of altitude. The MAORY optics will be more stable than those of the ELT being placed on the Nasmyth platform in a gravity invariant configuration. As discussed more in detail by [8], the MAORY optics positioning tolerances are:

$$(\Delta x, \Delta y, \Delta z) \rightarrow \pm 50 \ \mu m$$
$$(\Delta \theta x, \Delta \theta y) \rightarrow \pm 0.001 \ deg \quad (4)$$

In addition to the tolerances in (4) the MAORY bench is expected to expand/contract for temperature variations and to slightly torque with telescope tracking and pointing variations. Only the linear thermal variations are taken into account in the current analysis. The current strategy to achieve this level of positioning accuracy is applying a passive thermal control to maintain aligned the MAORY optics within the tolerances, by exploiting the thermal inertia (to minimize the thermal gradient over the night of the bench and optics) and trying to get an optical bench almost isothermal.
In this study the two DMs are kept in their nominal shape.

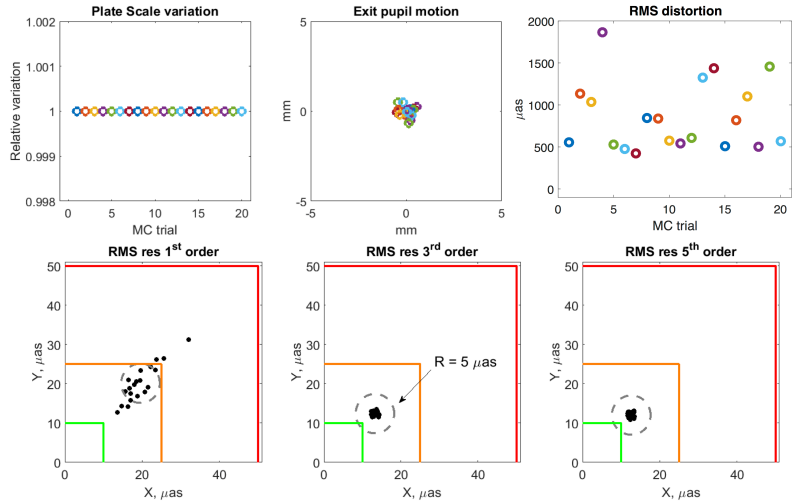

**Figure 4 Top**: PS variation, exit pupil motion and RMS distortion over the MICADO FoV induced by the MAORY optics positioning tolerances. **Bottom:** RMS residual (post-fit) over the MICADO FoV between the $n^{th}$ MC grid distortion and the nominal one after tolerancing. As for ELT, the $3^{rd}$ fit removes the majority of the distortion and the $5^{th}$ fit does not bring any significant further improvement (neither RMS lowering nor increased residuals compactness).

The MAORY tolerances have much smaller impact in terms of PS variations and exit pupil motion wrt the perturbations induced by ELT as well as the RMS distortions pre-fit that are about an order of magnitude smaller. The RMS distortions have a wide spread, about an order of magnitude, between different MC realizations. Like for the ELT, the $5^{th}$ order polynomial fit does not produce further improvement in the fit of the distortions.

## 3.3 MICADO tolerances and distortion

The MICADO optics are immersed in a cryogenic environment at about 40 K. As discussed in [1] the MICADO optical design is based on two TMAs mounted on fixed supports with no moving parts. Typical cryogenic integration tolerances for the TMAs are given by [9]:

$$(\Delta x, \Delta y, \Delta z) \rightarrow \pm 0.5 \ \mu m$$
$$(\Delta \theta x, \Delta \theta y) \rightarrow \pm 0.001 \ deg \quad (5)$$

The intrinsic distortion of the TMAs is significant (max ~ 2.5 % equivalent to 1.3" over FoV) over the instrument FoV, and it can vary as a consequence of the thermal gradients and the opto-mechanical tolerances. Currently non-conclusive numbers for the thermal stability of the cryostat are available, but the experience from previous instruments indicates a stability of ~ 0.1K/h with passive thermal control (as envisaged for MICADO) [10]. In addition, the distortion pattern from MICADO will rotate wrt the MAORY pattern as the instrument is derotated. The intrinsic distortions of MICADO and MAORY and their relative derotation-dependency shall be calibrated with the help of two astrometric masks, the first placed at the MAORY entrance FP and the second at the MICADO entrance FP [11]. The astrometric masks will provide a dense array of artificial point-like sources of known position and suitable SNR leading to efficiently and univocally map the internal distortions of the instruments. As shown in Figure 5, a denser grid of points (30x30) typical of an astrometric mask with ~1000 pinholes leads to smaller RMS residuals and a smaller spread between different MC realizations.

Also the MICADO tolerances have small impact in terms of PS variations and exit pupil motion wrt the perturbations induced by ELT while the RMS distortion pre-fit are at an intermediary level between the ELT and the MAORY ones. The RMS distortions between different MC realizations have a smaller spread wrt the ELT and MAORY case.

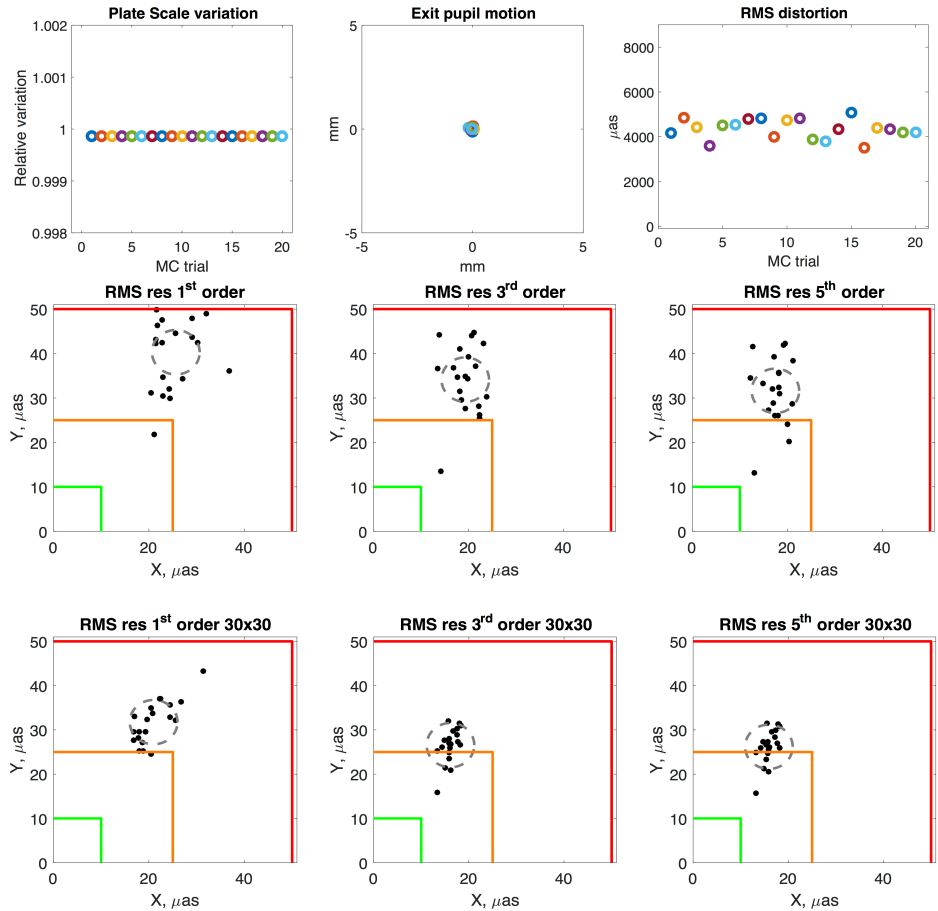

**Figure 5 Top**: PS variation, exit pupil motion and RMS distortion over the FoV induced by the MICADO optics positioning tolerances. **Centre:** RMS residuals (post-fit) over the MICADO FoV after tolerancing for a 12x12 grid of sampling points. **Bottom:** the RMS residuals (post-fit) for a denser grid of points (30x30) clusters in a more compact cloud with a smaller

barycentre wrt the 12x12 standard grid (Centre). A grid of 30x30 points mimics the astrometric mask functionality sampling efficiently the distortion pattern of the instrument down to the highest spatial frequencies.

## 4. PRELIMINARY ASTROMETRIC ERROR BUDGET & WORST OFFENDERS

The intrinsic expected maximum distortions (at the corner of MICADO FoV) are respectively: 0.0021% for ELT, 0.13 % for MAORY and 2.5 % for MICADO. MAORY and MICADO have also spatially more complex and anisotropic distortion patterns as shown in Figure 6. The preliminary astrometric error budget analysis reported in this work is however mostly oriented to the impact on the observations of the time-varying tolerances. The error budget (Figure 7, left) shows the predominance of the ELT distortions induced by the M2 tolerances. The MICADO TMAs are also very sensitive to tolerances, but the pie chart has to be read in relation with the typical timescale variations of the systems (Figure 7, right). The typical ELT timescales for distortion variation is 5 min and it is represented by the so-called Low Order Optimization (LOO) that recollimates M2 to M1 against the gravity flexure drift. The LOO occurs every 5 minutes with typical repositioning input steps of 50-100 μm [5]. The MAORY and MICADO timescales are expected to be longer, ~ hour, and related mainly to the night-time thermal gradient for the former and the internal cryostat thermal gradients for the latter.

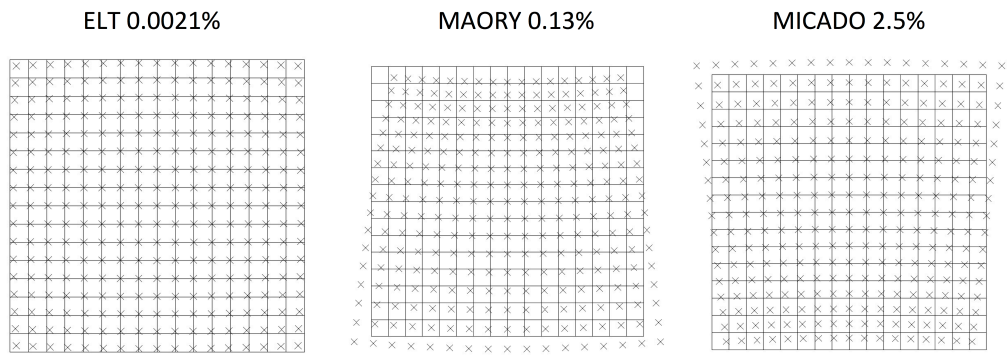

**Figure 6**: Focal plane distortion over the MICADO FoV (53") of ELT, MAORY and MICADO. The ELT distortion has rotational symmetry while those of MICADO and MAORY are anisotropic. The magnitude of the distortion grid has been rescaled to help visualization.

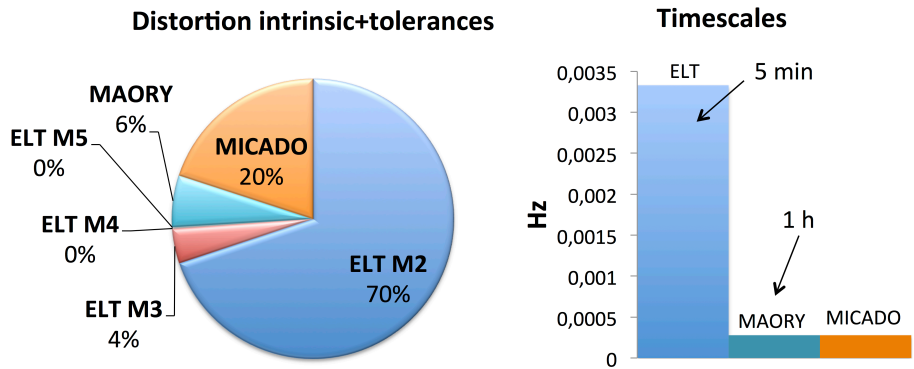

**Figure 7 Left**: Pie chart of the intrinsic distortions plus the distortion due to the tolerances for the ELT mirrors, MAORY and MICADO; the worst offender is M2 from ELT. **Right:** Typical timescales of variation of the distortion pattern for ELT (LOO timescales), MAORY night-time gradient 0.6-1K/h and MICADO 0.1K/h within the cryostat.

Looking at the RMS residuals pie chart after the polynomial fit (Figure 8) the ELT distortion contributions reduce significantly already with the 1st polynomial fit while MICADO and MAORY require the 3rd polynomial fit to decrease the absolute magnitude of the RMS residual below 50 µas.

As seen in Section 3 and in Figures 7 and 8, the ELT dominant distortions are PS variations over minutes timescales. This potentially limits the maximum MICADO astrometric exposure times and it requires a calibration strategy. The PS drift induced by ELT-M2 is in the order of ~ 35mas/30"/5min, while the optimum for MAORY-MICADO would be ~ 0.5-1mas/30". The ΔPS is mostly due to Δz ~ 0.1 mm of ELT-M2, and as as far as we know, ELT cannot internally correct for this drift during the LOO to better than 0.1 mm.

One possible solution could be using the MAORY post focal DMs to compensate for a PS variation induced by ELT and restore its nominal value provided that this operation do not introduce significant additional distortions. To compensate for a ΔPS ~ 1% induced by ELT a stroke of ~ 20-30 µm should be applied to the MAORY DMs (see Figure 9). A delicate point of the astrometry error budget control strategy in this case is the assessment the level of distortion introduced by this correction from MAORY in comparison with other options internal to the ELT like the modification of the M3 piston. The ELT tertiary mirror in fact is a crucial component of the telescope control strategy and performs a critical function in permitting the telescope to achieve a variable focal length [6]. Also this latter case needs to be studied both for feasibility at a telescope control strategy level and for the additional distortion introduced.

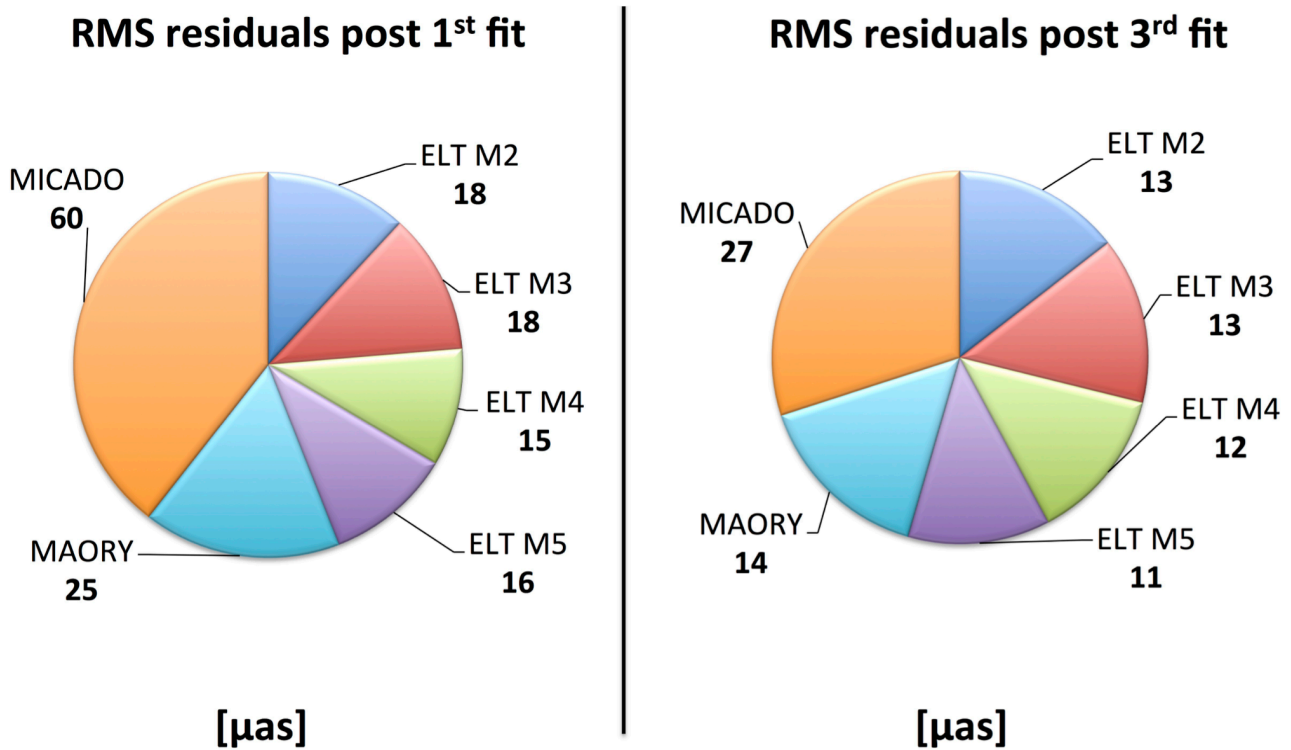

**Figure 8 Left**: Pie chart of the residual RMS distortion after 1st polynomial fit: the distortion from the ELT optics are more limited wrt to Figure 7 meaning that they mostly cause PS errors. The residual from MICADO are still large and require 3rd – 5th order polynomial fit to decrease below 50 µas as shown in the pie chart on the **Right.**

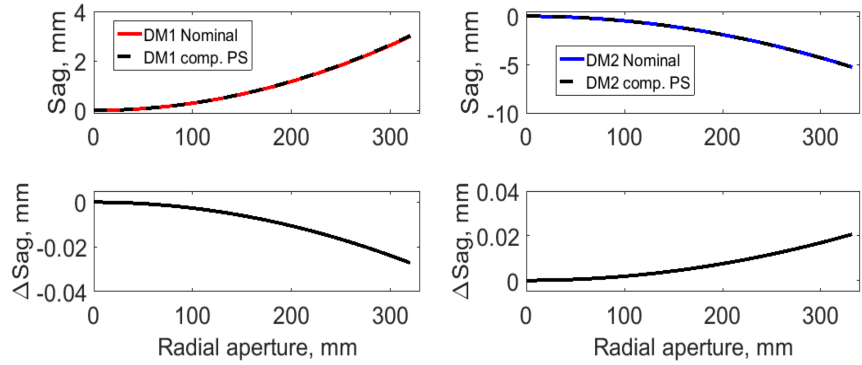

**Figure 9** MAORY DMs required stroke for compensating a ΔPS ~ 1 % coming from ELT. DM1 (left) DM2 (right). A PS error can be compensated also modifying the piston of M3 in ELT.

## CONCLUSIONS

This preliminary astrometric error budget for MICADO has been carried out with a hybrid approach, ray-tracing combined with MC tolerance studies and post-processing polynomial fit. The precision and reliability of the fitting routine has been verified with other two independent standard polynomial fit routines [13] [14], and one fitting tool based on Chebyshev polynomials [15]. All the fit routines show a good agreement on the RMS residuals both for ELT and MICADO grid of points; the Chebyshev polynomials lead to systematically lower RMS residuals (~10%) w.r.t. all the other fit routines. The astrometric error budget has led to a first sensitivity analysis from which the ELT-M2 is the worst offender. The typical axial displacement of M2 wrt M1 causes significant (0.1 %) PS variation over a few minutes timescales that in practice limits the MICADO exposure time to max 120 s. Over minutes timescales the PS variations smear out the PSF at the edge of the MICADO FoV challenging the astrometric requirements. The intrinsic distortions from MAORY and MICADO are relatively bigger than those from ELT, but much more stable in time against typical tolerances; the typical variation timescales of the distortion pattern follows the night-time thermal gradient ~ 0.6-1K/h for MAORY and the cryostat internal thermal variations ~ 0.1K/h for MICADO. The ultimate calibration of the telescope distortions has to be done on sky, but the study shows how these are well fitted already by a $3^{rd}$ polynomial, limiting the number of required field stars to 10-20. The calibration of MICADO and MAORY distortions relay instead on two astrometric masks deployable at their entrance focal planes and sampling the distortion pattern down to its highest spatial frequencies. Alternative/complementary strategies to enable 50 μas differential astrometry could involve a PS control strategy beyond the ELT LOO loop with the MAORY DMs or the ELT-M3. Limiting the instrument FoV over which performing astrometric studies and/or integrating faster (at the expenses of a higher RON) are also alternative scenarios to be investigated.